# Observation of soliton pulse compression in photonic crystal waveguides

P. Colman[1,3,*], C. Husko[1,2,*], S. Combrié[1,*], I. Sagnes[3], C. W. Wong[†,2], and A. De Rossi[†,1]

[1] *Thales Research and Technology, Route Départementale 128, 91767 Palaiseau, France*
[2] *Optical Nanostructures Laboratory, Center for Integrated Science and Engineering, Solid-State Science and Engineering, and Mechanical Engineering, Columbia University, New York, NY 10027*
[3] *Laboratoire de Photonique et de Nanostructures (CNRS UPR 20), Route de Nozay, 91460 Marcoussis, France*

[*] *Equal contribution*
[†] e-mail: alfredo.derossi@thalesgroup.com ; cww2104@columbia.edu

Soliton-effect pulse compression and propagation has been experimentally demonstrated in the temporal domain in fibers [1], fiber Bragg gratings [2], photonic crystal fibers (PhCF) [3-6], photonic nanowires [7] and in the spectral domain of integrated channel waveguides [8, 9]. Here we demonstrate the first experimental observations of soliton-effect pulse compression in semiconductor photonic crystal waveguides (PhCWG) in the temporal domain. Pulse compression in the PhCWGs occurs due to the interaction between a strong group-velocity dispersion (GVD) [10] and slow-light enhanced self-phase modulation (SPM) in the periodic dielectric media [11]. Compression of 3 ps input pulses to a minimum pulse duration of 580 fs (~10 pJ) is achieved. The small modal area $A_{eff} \sim 10^{-13}$ m$^2$ combined with a slow-light enhanced optical field allow for ultra-low threshold (~GW/cm$^2$) pulse compression at millimeter length scales. These results open the way for femtosecond and soliton applications on the chip scale.

Periodic dielectric structures have long been known to have extremely large dispersion thus enabling observation of soliton effects at centimeter length scales [2, 12-14]. The decreased interaction length, however, requires a correspondingly larger intensity-dependent nonlinear effect to match the strength of the dispersion. Increasing the optical intensity inside the waveguide is accomplished through: (a) direct input of larger peak powers; (b) decreasing the effective modal area [7-9]; or, most recently, (c) via dispersion-engineered periodic slow-light structures [10, 11]. At the so-called slow-light frequencies of PhCWGs, the light experiences a longer effective path length through the lattice via multiple Bragg reflections, leading to an enhanced local field density. The enhanced field scales inversely with the group velocity, thus decreasing the threshold of nonlinear effects such as Kerr, multi-photon absorption, or Raman scattering [15, 16]. In addition to the aforementioned *dispersion* engineering, additional research has centered on reducing the required intensity of chip-scale nonlinear effects via *material* engineering. Semiconductors [11, 17], chalcogenides [18-20], tellurites [21], and doped silica [22], have Kerr coefficients that are typically at least two orders of magnitude larger than silica fiber, further reducing the required energy threshold in integrated structures. In summary, the highly confined optical modes in dispersion-engineered semiconductor PhCWGs offer a long awaited development of low-intensity threshold nonlinearities to compliment the well-established large dispersion of periodic media [12-14, 23, 24].

Our examinations are performed in a GaInP photonic crystal membrane with a hexagonal lattice constant $a$ of 485 nm, $0.20a$ hole radius and a 170 nm thickness, that has an added line-defect of dielectric – or termed a photonic crystal waveguide [10, 25] – as shown in Figure 1a. The dispersion is tuned by increasing the innermost hole radii to $0.22a$. The 1.3-mm PhC waveguide supports two propagation modes, and includes integrated mode-adapters [26] to reduce the total input-output chip insertion losses to ~ 8 dB and suppress facet Fabry-Perot oscillations. The propagation loss, measured at $\alpha = 1$ dB/mm at $n_g = 6$, scales with $n_g^2$ [27]. The waveguide dispersion properties are measured via the phase-shift method, a method entailing the measurement of the radio-frequency phase modulation of the optical carrier, while sweeping the optical frequency through the waveguide transmission [28]. The measured phase shift $\Delta\phi$ is translated into a time delay through the relationship $\Delta T = \Delta\phi / 2\pi f$ with $f$ the optical carrier frequency and $\Delta T$ the propagation delay. The group index is thus $n_g = c\Delta T / L$. Figure 1b shows the group index $n_g$ increases from 5 to about 13 in the range of interest. The group-velocity dispersion (GVD, $\beta_2$) and third-order dispersion ($\beta_3$) coefficients, plot in Figure 1c, are obtained from fitting function numerical derivatives of the group index. The group velocity dispersion is negative (anomalous dispersion) and on the order of $\sim$ps$^2$/mm across the range of interest.

For the pulse compression experiments we employed a mode-locked fiber laser (Keopsys/Pritel) delivering nearly transform-limited pulses with full-width at half-maximum (FWHM) of 2.5 ps to 4 ps at a 22 MHz repetition rate. The source is tunable from 1525 to 1565 nm. After adjusting the source to the desired wavelength, we measured the pulse duration and minimized the time-band product to approach the Fourier-limit of hyperbolic secant pulses ($\Delta\lambda \Delta\nu = 0.315$) within 5% (i.e. negligible chirp). The pulse power is modulated with a half-wave plate and polarizer, thereby preventing misalignment and undesirable modification of the pulse shape. Autocorrelation traces were recorded using a PulseCheck APE autocorrelator directly coupled to the output of the waveguide. Importantly, we did not use any amplification in order to prevent artifacts and pulse distortion. We averaged over 256 autocorrelation traces, thereby improving the signal-to-noise ratio.

The nonlinear and dispersive effects for the optical solitons are captured by two lengthscales [29], a nonlinear length $L_{NL}$ ($= \frac{1}{\gamma P_o}$, where the nonlinear parameter $\gamma_{eff} = \frac{n_2 \omega_0}{c A_{eff}} \cdot \left(\frac{n_g}{n_o}\right)^2$ [13] and $P_o$ is the pulse peak power) and a dispersion length $L_D$ ($= \frac{T_o^2}{|\beta_2|}$, where $T_o = T/\Gamma$, $T$ is the pulse width (FWHM) and $\Gamma = 2\sinh^{-1}(\sqrt{2}) = 1.76$). For 3 ps pulse widths, the dispersion length ranges from 6 mm to 1.6 mm. The PhC waveguide chips are designed for the 1.3-mm physical length $L$ to be comparable to $L_D$, in order to support the optical solitons. The nonlinear Kerr parameter $\gamma_{eff}$ is inferred from the spectral broadening dependence on pulse peak power [30]. With increasing input pulse energies, the output pulse demonstrates the self-phase modulation broadened spectra with a π-phase shift at 7.3 pJ pulse energies (~ 2.5 W peak power) and group indices of 9.3 at 1555 nm. The measured effective nonlinear parameter $\gamma_{eff}$ is strongly dependent on $n_g$ (Figure 1d), with the largest value is just above 900 W$^{-1}$m$^{-1}$ at 1555 nm ($n_g$ of 9.3). At larger group indices ($n_g$ greater than 10) disorder-induced scattering [31] and three-photon absorption [30] begins to limit the effective Kerr nonlinearity and the measured $\gamma_{eff}$ parameter deviates from theoretical values. The GaInP material selection completely suppresses any two-photon absorption due to the large 1.9-eV band gap, well above the energy threshold for two-photon absorption (1.6-eV) such that residual effects from band tail absorption is negligible, as detailed in our recent studies [32]. From fibers to integrated structures, Table 1 shows the $\gamma_{eff}$ and $\beta_2$ parameters of various systems in which soliton-effects have been observed. Comparing briefly the current system to the 'record-holders' in each category, the observed $\gamma_{eff}$ is on the same order of magnitude as integrated nanowires, though slightly larger [8, 9]. Regarding the GVD, the present system is a scant factor of two smaller than fiber Bragg gratings (FBG). However, the semiconductor nanowires have a GVD that is ~10$^2$ smaller, while the FBG, have a $\gamma_{eff}$ that is ~10$^5$ less in magnitude, and consequently require much larger input powers (~kW). The simultaneous combination of both large $\gamma_{eff}$ and GVD in a single material system, along with suppressed TPA, enable compression at pJ energies and millimeter lengthscales in the GaInP PhCWG.

We next examined the output pulses directly in the time-domain through second-harmonic intensity autocorrelation. Figure 2a show a series of traces at 1551 nm ($n_g$ of 8.3) for increasing pulse energies. The autocorrelation trace widths (FWHM) decrease from an input duration of 4.9 ps to a minimum 900 fs at 22 pJ, for a compression ratio, $\chi_c$ (= $T_{in}/T_{out}$), of 5.4. Employing an autocorrelation deconvolution factor of 1.54 for hyperbolic secant pulses (direct extraction performed later in this work and also matches exactly), this implies that the pulse width is reduced from 3.2 ps to 580 fs. Increased spectral broadening with pulse energy due to SPM collected via OSA is shown in the right panel of Figure 2a.

The propagation of optical pulses in a slow-light PhC waveguide is modeled through the nonlinear Schrödinger equation (NLSE) [24, 29]. In contrast to gap solitons, which are approximately described as superposition of Bloch modes at the band edge [23, 33], not reachable here due to large backscattering [31]. Under these conditions the PhC behaves as a single-mode waveguide, where dispersion and nonlinearity are referred to Bloch modes [34] in lieu of translation-invariant modes. The NLSE model is described by:

$$\frac{\partial E}{\partial z} = -\frac{1}{v_g}\frac{\partial E}{\partial t} - i\frac{\beta_2}{2}\frac{\partial^2 E}{\partial t^2} + \frac{\beta_3}{6}\frac{\partial^3 E}{\partial t^3} - \frac{\alpha_1}{2}E + i\gamma_{eff}|E|^2 E - \frac{\alpha_{3eff}}{2}|E|^4 E + (ik_o\delta - \frac{\sigma}{2})N_c E \ . \qquad (1)$$

This includes third-order dispersion $\beta_3$, linear propagation loss $\alpha$, effective slow-light three-photon nonlinear absorption $\alpha_{3eff}$ [30], effective nonlinear parameter $\gamma_{eff}$, and generated carrier density $N_c$ with associated free-carrier dispersion $\delta$ and absorption $\sigma$. The auxiliary carrier equation introduces a non-instantaneous response through the carrier lifetime $\tau_c$:

$$\frac{\partial N_c}{\partial t} = \frac{\alpha_{3eff}}{3\hbar\omega A_{3eff}}|E|^6 - \frac{N_c}{\tau_c} \ .$$ The free-carrier dispersion coefficient $\delta$ includes $n_g$ scaling: . Here $\sigma$ is $4\times10^{-21}(n_g/n_0)$ m$^2$ based on literature, with bulk index $n_0$. We solve the NLSE model employing an implicit Crank-Nicolson split-step method. Parameters are obtained directly from experimental measurements or calculated as required ($A_{3eff}$) [35]. Third-order dispersion, included in the model, contributes negligibly throughout the range of parameters examined here, even at the minimum pulse durations.

The measured autocorrelation (black dots) with corresponding numerical traces (red solid) from the NLSE at 1551 nm and 1555 nm are shown in Figs. 2(b) and (c), respectively. The fit to an ideal hyperbolic secant pulse shape (blue dashed), and the compression factor $\chi_c = T_{in}/T_{out}$ is also shown. Several corresponding spectral traces and simulations at 1555 nm are shown in Fig. 2(d). Importantly, the model agrees simultaneously with both the autocorrelation and spectra with no degrees of freedom. The nearly symmetrical spectra indicate free-carrier effects play a minor role, except at the largest pulse energies and group indices. The main impact of the dissipative terms in the NLSE is to "slow" down the soliton dynamics, e.g. to increase the effective spatial scale. The ultimate limits to the system are two-fold. The strong linear scattering of slow-light ultimately eliminates the periodic property of higher-order solitons. As for the nonlinear absorption, ThPA places a fundamental limit to the peak powers that can be produced, and thus forms the upper bound to the maximum compression. We note that materials limited by two-photon absorption would experience greater attenuation and carrier dispersion, (e.g. have lower peak powers) and thus experience even greater limitations to compression. They also require longer length scales to observe the same phenomena, as they possess a greatly diminished critical intensity from free-carrier effects [10].

We additionally note that for an anomalous dispersion waveguide with positive chirp, temporal compression could possibly be observed in the initial lengths of the waveguide without self-phase modulation or soliton formation. We experimentally verified that this effect is negligible via a number of methods. First, as the input pulses are nearly transform-limited, pre-input chirp is very small. Second, at low power (less than 1 pJ), we observed that the output pulse width is identical (within measurement error) to the input pulse. Third, the compression is directly related to the increase of the coupled energy, controlled with a half-waveplate, such that the input pulse shape remains unmodified throughout the experiment. In addition, we note that the Raman contribution and related self-frequency shift [9, 16, 36] is negligible at our power levels for this material as seen in our pulse spectra measurements. Further increase of the compression factor relies on suppression of three-photon absorption with improved materials and nanofabrication, along with examinations of dispersion-managed PhCWGs for chirped or flat dispersion at low group velocities.

We have demonstrated pulse compression based on high-order solitons at moderately-slow group velocities in GaInP photonic crystal waveguides. This possibility is enabled by the enhanced self-phase modulation and strong negative group velocity dispersion in the photonic crystal waveguides. Use of a material free of two-photon absorption dramatically reduces the impact of nonlinear absorption and free-carrier dispersion, thus preventing detrimental interference with the soliton dynamics. The soliton dynamics emerge from temporal and spectral measurements and are further reinforced with a nonlinear Schrödinger model, leading to quantitative agreement with experiments. Owing to the small size of the device (1.3-mm) and low energies (~ 20 pJ), these results are promising developments towards the integration of femtosecond and soliton applications in photonic chips.

**References**

[1] L. F. Mollenauer, R. H. Stolen, and J. P. Gordon, Experimental Observation of Picosecond Pulse Narrowing and Solitons in Optical Fibers, *Phys. Rev. Lett.* **45**, 1095 (1980).

[2] B. Eggleton, R. Slusher, M. de Sterke, P. Krug, and J. Sipe, Bragg grating solitons, *Phys. Rev. Lett.* **76**, 1627, (1996).

[3] D. G. Ouzounov, F. R. Ahmad, D. Müller, N. Venkataraman, M. T. Gallagher, M. G. Thomas, J. Silcox, K. W. Koch, and A. L. Gaeta, Generation of Megawatt Optical Solitons in Hollow-Core Photonic Band-Gap Fibers, *Science* **301**, 1702 (2003).

[4] F. Gerôme, K. Cook, A. K. George, W. J. Wadsworth, and J. Knight, Delivery of sub-100fs pulses through 8m of hollow-core fiber using soliton compression, *Opt. Express* **15**, 7126 (2007)

[5] J. M. Dudley, J. R. Taylor, Ten years of nonlinear optics in photonic crystal fibre, *Nature Photonics* **3**, 85 (2009).

[6] A. A. Amorim, M. V. Tognetti, P. Oliveira, J. L. Silva, L. M. Bernardo, F. X. Kärtner, and H. M. Crespo, Sub-two-cycle pulses by soliton self-compression in highly nonlinear photonic crystal fibers, *Opt. Lett*. **34**, p. 3851 (2009).

[7] M. Foster, A. Gaeta, Q. Cao, and R. Trebino, Soliton-effect compression of supercontinuum to few-cycle durations in photonic nanowires, *Opt. Express* **13**, 6848 (2005).

[8] J. Zhang, Q. Lin, G. Piredda, R. W. Boyd, G. P. Agrawal, and P. M. Fauchet, Optical solitons in a silicon waveguide, *Opt. Express* **15**, 7682 (2007).

[9] J. I. Dadap, N. C. Panoiu, X. Chen, I-Wei Hsieh, X. Liu, C.-Y. Chou, E. Dulkeith, S. J. McNab, F. Xia, W. M. J. Green, L. Sekaric, Y. A. Vlasov, and R. M. Osgood, Jr, Nonlinear-optical phase modification in dispersion-engineered Si photonic wires, *Opt. Express* **16**, 1280 (2008).

[10] T. Baba, Slow light in photonic crystals, *Nature Photonics* **2**, 465 (2008).

[11] C. Monat, B. Corcoran, M. Ebnali-Heidari, C. Grillet, B.J. Eggleton, T. P. White, L. O'Faolain, and T. F. Krauss, Slow light enhancement of nonlinear effects in silicon engineered photonic crystal waveguides, Opt. Express **17**, 2944 (2009).

[12] H. G. Winful, Pulse compression in optical fiber filters, *App. Phys. Lett*. **46**, 527 (1985).

[13] R. E. Slusher and B. J. Eggleton, eds., *Nonlinear Photonic Crystals*, Springer Verlag, Berlin, Germany, 2003.

[14] Y. S. Kivshar and G. P. Agrawal, *Optical Solitons: From Fibers To Photonic Crystals*, Academic Press, San Diego, CA, 2003.

[15] M. Soljačić and J. D. Joannopoulos, Enhancement of nonlinear effects using photonic crystals, Nature Materials **3**, 211 (2004).

[16] J. F. McMillan, M. Yu, D.-L. Kwong, and C. W. Wong, Observations of spontaneous Raman scattering in silicon slow-light photonic crystal waveguides, *Appl. Phys. Lett.* **93**, 251105 (2008).

[17] K. Inoue, H. Oda, N. Ikeda, and K. Asakawa, Enhanced third-order nonlinear effects in slow-light photonic-crystal slab waveguides of line- defect, *Opt. Express* **17**, 7206 (2009).

[18] D. I. Yeom, E. Mägi, M. R. E. Lamont, M. A. F. Rolens, L. Fu, and B. J. Eggleton, Low-threshold supercontinuum generation in highly nonlinear chalcogenide nanowires, *Opt. Lett*. **33**, 660 (2008).

[19] M. R. E. Lamont, B. Luther-Davies, D.-Y. Choi, S. Madden, and B. J. Eggleton, Supercontinuum generation in dispersion engineered highly nonlinear ($\gamma = 10$ /W/m) $As_2S_3$ chalcogenide planar waveguide, *Opt. Express* **16**, 14938, (2008).

[20] K. Suzuki, Y. Hamachi, and T. Baba, Fabrication and characterization of chalcogenide glass photonic crystal waveguides, *Opt. Express* **17**, 22393, (2009).

[21] M. Liao, C. Chaudhari, G.i Qin, Xin Yan, T. Suzuki, and Y. Ohishi, Tellurite microstructure fibers with small hexagonal core for supercontinuum generation, *Opt. Express* **17**, 12174 (2009).

[22] D. Duchesne, M. Ferrera, L. Razzari, R. Morandotti, B. E. Little, S. T. Chu, and D. J. Moss, Efficient self-phase modulation in low loss, high index doped silica glass integrated waveguides, *Opt. Express* **17**, 1865-1870 (2009)

[23] W. Chen and D. L Mills, Gap solitons and the nonlinear optical response of superlattices, *Phys. Rev. Lett.* **58**, 160 (1987).

[24] J. E. Sipe and H. G. Winful, Nonlinear Schrödinger solitons in a periodic structure, *Opt. Lett.* **13**, 132 (1988).

[25] M. Notomi, K. Yamada, A. Shinya, J. Takahashi, C. Takahashi and I. Yokohama, Extremely large group-velocity dispersion of line-defect waveguides in photonic crystal slabs, *Phys. Rev. Lett.* **87**, 253902 (2001).

[26] Q. Tran, S. Combrié, P. Colman, and A. De Rossi, Photonic crystal membrane waveguides with low insertion losses, *Appl. Phys. Lett.* **95**, 061105 (2009).

[27] R. Engelen, D. Mori, T. Baba, L. Kuipers, Two regimes of slow-light losses revealed by adiabatic reduction of group velocity, *Phys. Rev. Lett.* **101**, 103901 (2008).

[28] S. Combrié, A. De Rossi, L. Morvan, S. Tonda, S. Cassette, D. Dolfi and A. Talneau, Time-delay measurement in singlemode, low-loss photonic crystal waveguides, *Electron. Lett.* **42**, 86 (2006).

[29] G. P. Agrawal, *Nonlinear Fiber Optics*, Academic Press, San Diego, CA, 2007.

[30] C. Husko, S. Combrié, Q. V. Tran, F. Raineri, C. W. Wong, and A. De Rossi, Non-trivial scaling of self-phase modulation and three-photon absorption in III-V photonic crystal waveguides, *Opt. Express* **17**, 22442 (2009).

[31] M. Patterson, S. Hughes, S. Combie, N.-V. Quynh Tran, A. De Rossi, R. Gabet and Y. Jaouen, Disorder induced coherent-scattering in slow light photonic crystal waveguides, *Phys. Rev. Lett.* **102**, 253903 (2009).

[32] S. Combrié, Q. V. Tran, C. Husko, P. Colman, and A. De Rossi, High quality GaInP nonlinear photonic crystals with minimized nonlinear absorption, *Appl. Phys. Lett.* **95**, 221108 (2009).

[33] A. De Rossi, C. Conti, and S. Trillo, Stability, Multistability, and Wobbling of Optical Gap Solitons, *Phys. Rev. Lett.* **81**, 85 (1998).

[34] N. A. R. Bhat and J. E. Sipe, Optical pulse propagation in nonlinear photonic crystals, *Phys. Rev. E* **64**, 056604 (2001).

[35] S. G. Johnson and J. D. Joannopoulos, Block-iterative frequency-domain methods for Maxwell's equations in a planewave basis, *Opt. Express* **8**, 173 (2001).

[36] R. Claps, D. Dimitropoulos, Y. Han, and B. Jalali, Observation of Raman emission in silicon waveguides at 1.54 mm, *Opt. Express* **10**, 1305 (2002).

**Acknowledgments**

This work is supported in part by the European Commission GOSPEL project (grant n. 219299) and the French Research Agency project L2CP (S.C., P.C., and A.D.R.), the National Science Foundation CAREER Award (0747787) and ECCS (0725707) (C.A.H. and C.W.W.), the Fulbright Foundation (C.A.H.), and the New York State Foundation for Science, Technology and Innovation (C.W.W.). The authors acknowledge valuable discussions with Profs. Erich Ippen, Franz Kaertner, and Stefano Trillo. The authors thank Quynh-Vy Tran for his contributions to the development of the PhC technology in Thales and Alexandre Chen (Alcatel-Thales III-V Lab).

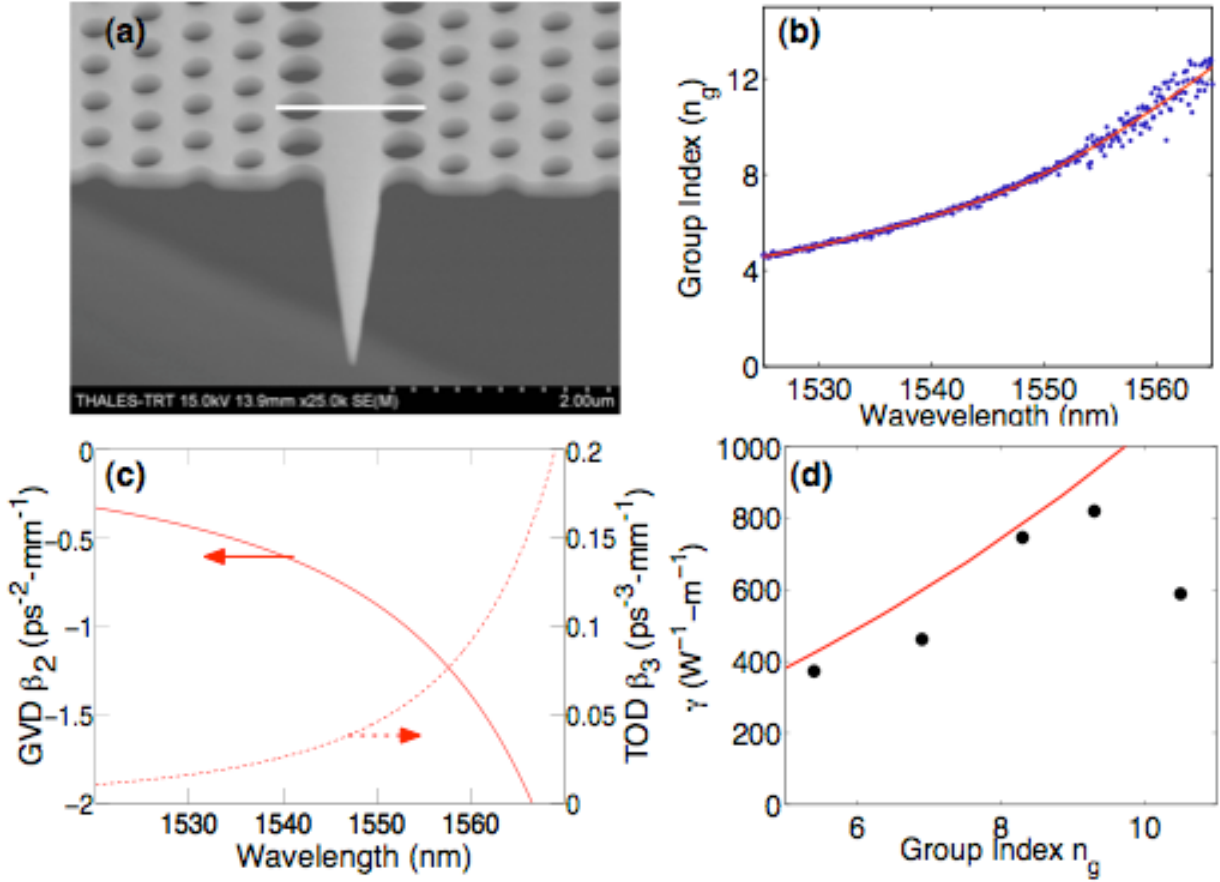

**Figure 1** | Dispersion and slow-light properties of the GaInP photonic crystal waveguide sample. (a) Scanning electron micrograph of GaInP membrane with designed mode adapters [25] SEM image of the GaInP PhCWG sample (Scale: 1um)  (b) Group index measured (markers) and fit (solid line) [27] (c) Group velocity dispersion (left axis) and third order dispersion (right axis) derived from phase-shift group index measurements [27]. At 1551nm the GVD is -0.91 ps$^2$/mm (d) Nonlinear parameter, $\gamma_{eff}$ extracted from experimental (dots) and theoretical scaling (line) [14] due to slow-light enhancement.

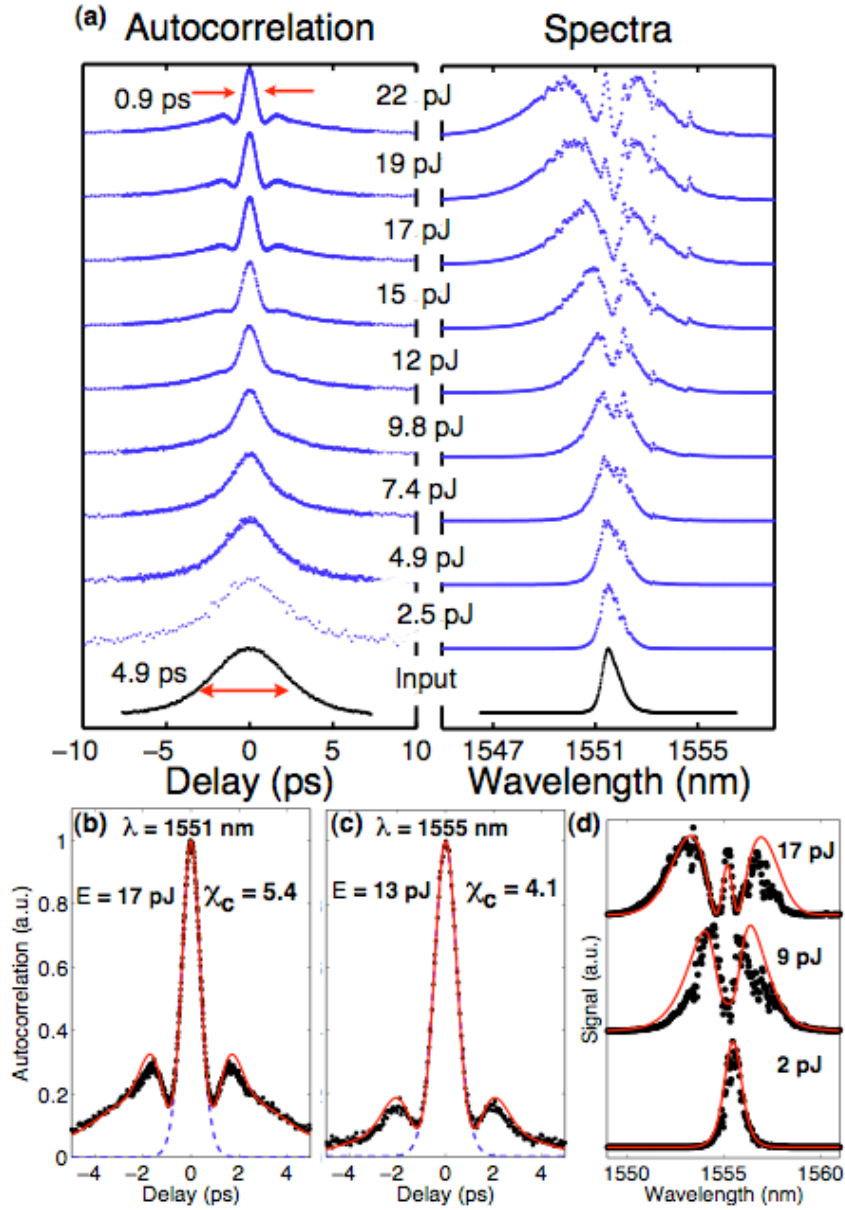

**Figure 2** | Pulse autocorrelation and spectra measurements. (a) Experimental intensity autocorrelation and corresponding spectra of output pulses for increasing coupled pulse energies from 1.7 pJ to 20 pJ and corresponding spectra for pulses centered at 1551 nm ($n_g$=8.3). A minimum pulse width of 580 fs (sech$^2$ deconvolved; 1.54 factor) is achieved. Measured autocorrelation (black dots) at (b) 1551 nm and (c) 1555 nm. Corresponding numerical traces (red solid) from NLSE and fit to an ideal hyperbolic secant (blue dashed), and the compression factor $\chi_c = T_{in}/T_{out}$ is also shown. (d) Spectrum traces and simulation at 1555 nm.
11

**Table 1** | Dispersion and nonlinear parameters of different material systems with experimentally demonstrated soliton effects

$L_D$ with $T_0 = 2$ ps    and    $L_{NL}$ with $P_{peak} = 1$ W

| System | $\beta_2$ [ps$^2$/m] | $\gamma_{eff}$ [W$^{-1}$m$^{-1}$] | $L_D$ [m] | $L_{NL}$ [m] | Ref. |
|---|---|---|---|---|---|
| Fiber | 2.20 x 10$^{-2}$ | 1.1 x 10$^{-3}$ | 182 | 909 | [29] |
| FBG | 2000 | 9.4 x 10$^{-3}$ | 2.0 x 10$^{-3}$ | 107 | [2, 13] |
| Hollow PhCF | 1.83 x 10$^{-2}$ | 2.1 x 10$^{-6}$ | 219 | 4.8 x 10$^5$ | [3] |
| NL PhCF | 2.00 x 10$^{-2}$ | 0.21 | 200 | 4.67 | [6] |
| Tellurite PhCF | 0.185 | 5.70 | 21.6 | 0.18 | [21] |
| Tapered PhCF fiber | 0.10 | 0.37 | 39.3 | 2.70 | [7] |
| Chalco tapered fiber | 0.36 | 93 | 11.1 | 1.1 x 10$^{-2}$ | [18] |
| Silica nanowire | 2.50 x 10$^{-2}$ | 0.2 | 160 | 5.00 | [22] |
| Si nanowire | 2.26 | 300 | 1.8 | 3.3 x 10$^{-3}$ | [8, 9] |
| Chalcogenide nanowire | 3.70 x 10$^{-2}$ | 10 | 108 | 0.10 | [19] |
| GaInP PhCWG | 1100 | 920 | 3.6 x 10$^{-3}$ | 1.1 x 10$^{-3}$ | This work $n_g$=8.3 |